\documentclass[11pt]{article}
\usepackage{amsmath}
\usepackage{amscd}
\newcommand{\ket}[1]{\left | #1 \right \rangle}
\newcommand{\bra}[1]{\left \langle #1 \right |}
\newcommand{\onarrow}[1]{\stackrel{#1}{\longrightarrow}}
\newcommand{\basis}[1]{{\cal B}_{#1}}

\newcounter{thm}
\newenvironment{theorem_env}{\textbf{Theorem
\arabic{thm}:}\begin{itshape}
}{\end{itshape}\addtocounter{thm}{1}}

\def\openone{\leavevmode\hbox{\small1\kern-3.8pt\normalsize1}}
\def\RR{{\rm I\kern-.2emR}}
\def\tr{{\rm tr}\; }
\def\ce{{\cal E}}

\def\ch{{\cal H}}

\def\cq{{\cal Q}}

\newcommand{\proj}[1]{\ket{#1}\!\bra{#1}}
\newcommand{\outerp}[2]{\ket{#1}\!\bra{#2}}
\newcommand{\ketbra}[2]{\outerp{#1}{#2}}
\newcommand{\tensor}{\otimes}

\newcommand{\braket}[2]{\langle #1|#2 \rangle}

\newcommand{\beqa}{\begin{eqnarray}}
\newcommand{\eeqa}{\end{eqnarray}}
\newcommand{\QED}{\hspace*{\fill}\mbox{\rule[0pt]{1.5ex}{1.5ex}}}

\begin{document}
\begin{center}
{\LARGE\bf Universal quantum information compression\\ and degrees
of prior knowledge \\ }
\bigskip
{\normalsize Richard Jozsa  and Stuart Presnell }\\
{\small\it Department of Computer Science, University of
Bristol,\\ Merchant Venturers Building, Bristol BS8 1UB U.K.}
\\[4mm]
\date{today}
\end{center}

\begin{abstract}
We describe a universal information compression scheme
that compresses any pure quantum i.i.d. source asymptotically to
its von Neumann entropy, with no prior knowledge of the structure 
of the source. 
We introduce a diagonalisation procedure
that enables any classical compression algorithm to be utilised 
in a quantum context. Our scheme is then based on the corresponding 
quantum translation of the classical Lempel--Ziv algorithm.
Our methods lead to a conceptually simple
way of estimating the entropy of a source in terms of the measurement
of an associated length parameter 
while maintaining high fidelity for long blocks.  
As a by-product we also estimate the 
eigenbasis of the source.
Since our scheme is based on the Lempel--Ziv method, it can be 
applied also to target sequences that are not i.i.d.

\end{abstract}
\section{Introduction}

In addition to its evident utility in practical communication issues,
the concept of information compression provides a bridge between the 
abstract theory of information and concrete physics -- it characterises 
the minimal physical resources (of various appropriate kinds) that 
are necessary and sufficient to faithfully encode or represent information.

In the case of quantum information, the study of optimal compression
rates is especially interesting as it relates directly to non-orthogonality 
of states \cite{js00} and entanglement, and thus provides a new tool for investigating
 foundational properties
of these uniquely quantum features. Almost all work to date on quantum 
information compression has studied compression properties of a so-called
independent identically distributed source (i.i.d. source). (However see 
 \cite{petz} for an interesting non-i.i.d. situation.) Let $\ce = \{
\ket{\sigma_i}; p_i \}$ be an ensemble of (pure) quantum signal states 
$\ket{\sigma_i}$ with assigned probabilities $p_i$. An i.i.d. source 
comprises an unending sequence of states chosen independently from $\ce$.
For each integer $n$ we have an ensemble of signal blocks of length $n$.
Writing $I= i_1 \ldots i_n$ the states are $\ket{\sigma_I}= 
\ket{\sigma_{i_1}}\otimes \ldots \otimes \ket{\sigma_{i_n}}$ with 
probabilities $p_I= p_{i_1}\ldots p_{i_n}$. 
Let $\ch$ (with dimension $d$) denote the Hilbert space of single signals and let $\cq_\alpha$
denote the space of all mixed states of $\alpha$ qubits (or the smallest 
integer greater than $\alpha$ if $\alpha$ is not an integer). Then $n$-blocks 
$\ket{\sigma_I}$ are in $\ch^{\otimes n}$ and in $\cq_{n\log d}$. 
(In this paper $\log$ will denote logarithms to base 2). To define the 
notion of compression we first introduce the fidelity 
\begin{equation} \label{fid}
F(\proj{\psi},\rho) = \bra{\psi}\rho \ket{\psi} \end{equation}
between any pure and mixed state. More generally if $\rho$ and $\omega$ 
are mixed we define fidelity by \cite{j94,u76}
\begin{equation} \label{genfid} F(\rho, \omega) = 
(\tr \sqrt{\sqrt{\omega}\rho\sqrt{\omega}})^2. \end{equation}
The von Neumann entropy $S$ of an ensemble $\ce$ is defined by 
\[ S=-\tr \rho \log \rho \]
where $\rho = \sum_i p_i \proj{\sigma_i}$ is the overall density 
matrix of the signals.

An encoding--decoding scheme for blocks of length $n$, to $\alpha$ 
qubits per signal and average fidelity $1-\epsilon$, is defined by 
the following ingredients: \\ (i) An encoding operation 
$E_n:\ch^{\otimes n} \rightarrow \cq_{n\alpha}$ which is a completely
positive trace preserving (CPTP) map. $E_n (\ket{\sigma_I})$ is a (mixed) 
state of $n\alpha$ qubits called the encoded or compressed version 
of $\ket{\sigma_I}$. \\ (ii) A decoding operation $D_n:\cq_{n\alpha} 
\rightarrow \cq_{n\log d}$ which is also a CPTP map. We write 
$\tilde{\sigma}_I=D_n E_n (\ket{\sigma_I})$ and call it the decoded 
version of $\ket{\sigma_I}$. Note that $\tilde{\sigma_I}$ is generally 
a mixed state. \\ (iii) The average fidelity between $\ket{\sigma_I}$
and $\tilde{\sigma}_I$ is $1-\epsilon$:
\[ \sum_I p_I F(\ket{\sigma_I},\tilde{\sigma}_I)=1-\epsilon \]
We say that the source $\ce$ may be compressed to $\alpha$ qubits per signal
if the following condition is satisfied: for all $\epsilon >0$ 
there is an $n_0$ such that for all blocks of length $n>n_0$ there 
is an encoding--decoding scheme for blocks of length $n$ to $\alpha$ qubits per signal
and average fidelity at least $1-\epsilon$.

The above definitions are motivated by source coding for i.i.d. sources 
in Shannon's classical information theory (cf \cite{covth} for an exposition).
Indeed if the signal states are mutually orthogonal and the coding/decoding 
operations are classical probabilistic processes, then we regain the standard 
classical theory. The quantum generalisation of Shannon's source coding theorem is
Schumacher's quantum source coding theorem \cite{Schu,js94,bcfj,w99}, 
stating that the optimal compression rate is the von Neumann entropy $S$ of the
signal ensemble. More precisely, if $\alpha \neq S$ then $\ce$ may be compressed 
to $\alpha$ qubits per signal iff $\alpha >S$. In these source coding theorems it is assumed
that we have knowledge of the signal ensemble states $\ket{\sigma_i}$ and 
their prior probabilities $p_i$. (Actually knowledge of the density matrix
$\rho=\sum_i p_i \proj{\sigma_i}$ alone suffices). 

The question of {\em universal} compression
concerns a situation in which we have only partial, or even no knowledge, about the
i.i.d. source $\ce$. We may even go further and ask about compressing a target sequence
from a source that is not even assumed to be i.i.d. (but perhaps has other properties 
e.g. a Markovian source). Thus universal compression may be studied 
in the presence of varying degrees of prior knowledge about the source.
In this paper we will consider universal compression of i.i.d. sources. However in 
contrast to all other quantum compression schemes proposed to date, our methods
can also be applied to non-i.i.d. sources in a natural way, in various situations
(that will become clear in our exposition below).

A classical i.i.d. source is fully characterised just by its probability distribution
$\{ p_i \}$ of signals. In a quantum i.i.d. source, for the purpose of 
studying the action of the encoding and decoding maps, each signal state may be 
taken to be in the mixed state $\rho=\sum_i p_i \proj{\sigma_i}$. 
Hence a quantum i.i.d. source is fully characterised by the classical 
probability distribution $\{ \lambda_i \}$ of the eigenvalues of $\rho$ {\em
together with} the specification of a corresponding orthonormal 
eigenbasis $\{ \ket{\lambda_i} \}$. The distribution $\{ \lambda_i \}$ is the direct
analogue of the distribution $\{ p_i \}$ of a classical source and the 
extra freedom in the quantum case, of the orientation of the eigenbasis makes the 
problem of universal quantum compression inherently more difficult than its 
classical counterpart.

Before presenting our main results we give a brief overview of existing work 
on universal quantum information compression. The basic technique of so-called 
Schumacher compression \cite{Schu,js94} used in the Schumacher source coding theorem utilises
the typical subspace of $\rho$. This construction requires knowledge of
both the eigenvalues and eigenvectors of $\rho$. As such, it does not appear 
to offer any generalisation to a universal compression scheme, 
with a prior knowledge of anything less than full knowledge of the source.
In \cite{jhhh} Jozsa et al. presented a universal compression scheme for 
quantum i.i.d. sources, requiring a prior knowledge of an upper bound 
$S_0>S$ on the von Neumann entropy of the source but requiring no prior
knowledge of the orientation of the eigenbasis of the source.
The scheme compressed the source to $S_0$ qubits per signal (in
contrast to the optimal $S$ in the case that the source is known).
Hence if the von Neumann entropy (or set of eigenvalues) of the source is known 
then this scheme is universal, with no prior knowledge of the eigenbasis.

In \cite{SW} and \cite{bfg} a quantum analogue of Huffman 
variable length coding was developed. The techniques of Schumacher 
and Westmoreland in \cite{SW} motivated the formulation of our two stage
compression model below. Although the Schumacher--Westmoreland scheme
is not presented as a universal scheme, if we adjoin the method of ``smearing''
measurements used by Hayashi and Matsumoto \cite{HM,HM2}  (and also described and
used by us below) then the scheme can be made universal for
 a situation in which the orientation of the eigenbasis is known
but the eigenvalue distribution is unknown.

The first fully universal compression scheme for quantum i.i.d. sources
was presented by Hayashi and Matsumoto \cite{HM}. This scheme compresses any 
quantum i.i.d. source to its von Neumann entropy $S$, requiring no prior 
knowledge of the eigenbasis or eigenvalues of the source density matrix.
Their method is based on the scheme of \cite{jhhh} supplemented by an 
estimation of the eigenvalues of the source and hence of $S$.
 
In this paper we present an alternative fully universal quantum compression scheme
with various novel features \cite{tokyo}. In classical information theory \cite{covth} 
there exists a variety of schemes for universal classical information compression. 
Some of these, such as the Lempez--Ziv method, apply even to situations in 
which the target string does not come from an i.i.d. source. Below we will introduce
a ``diagonalisation procedure'' which effectively enables any such classical scheme
to be transferred into the context of quantum compression. Then utilising the 
measurement smearing technique  of \cite{HM}, in conjunction with a further 
iterative procedure, we will achieve universal quantum information compression. 
In contrast to the scheme of \cite{HM} our scheme will include an estimation of 
the eigenbasis orientation, and we estimate the source entropy $S$ via a
conceptually simpler 
estimation of a block length parameter, whose knowledge is equivalent to that of $S$.
Our scheme will be based on transferring the classical Lempel--Ziv scheme 
into a quantum context. Hence (in contrast to any other existing scheme) 
it will be applicable even to sources that are not assumed to be i.i.d., although 
the question of optimality of the achieved compression rate in these more 
general situations remains to be explored.

\section{Two-Stage Compression Model}

Let $\ce = \{ \ket{\sigma_i}; p_i \}$ be the signal ensemble of a quantum i.i.d. 
source and let
$\rho = \sum_i p_i \proj{\sigma_i}$. In the signal state space let $ \{ \ket{e_i}\}$ 
be a fixed chosen basis called the computational basis.
The process of compressing blocks of length $n$ of the source will in
general consist of some unitary manipulations of the state
\(\rho^{\otimes n}\)
 (and we also include the possible adjoining of ancillary qubits in a standard state),
 and some operations which are not unitary (i.e. discarding qubits or 
measurement operations).
We know from the study of quantum circuits \cite{NC} that the
order of operations can be re-arranged to put all the non-unitary
steps at the end.  Thus the whole procedure can be naturally
divided into two separate stages.
The first, consisting of all the unitary manipulations of the
state, will
be called the ``condensation'' stage. This stage takes the form of
an algorithm to be performed by a quantum computer.  Naturally,
it cannot decrease the
total length of the sequence.

At the end of the condensation stage, the string should have been
manipulated in such a way that the first $n R_c$ qubits of output
contain a faithful representation of the data in the input (for
some ``condensation rate'' $S(\rho) \leq R_c \leq 1$).  The
remainder of the qubits are in a state asymptotically independent
of the input (for simplicity, we assume the state $\ket{0}$).  We
call these qubits ``blank''. 

The second stage consists of the measurement operations.  This is
where actual compression takes place, since now the dimension of
the Hilbert space in which the state lives can be reduced.  It is
called the ``truncation'' stage, since it tends to involve
removing the ``blank'' qubits at the end.  In truncation the
length of the string is reduced to \(n R_t\) for some $R_c \leq
R_t \leq 1$.  Asymptotic independence between the ``blanks'' and
the data qubits is equivalent to saying that this truncation can
be performed with fidelity $F\rightarrow1$ as
$n\rightarrow\infty$.

Determining exactly where the truncation cut is to be made is in
general a difficult task.  Previous compression schemes have
relied upon given  prior knowledge of the source to do this. 
In the present case we do not assume that such information is 
given {\em a priori}.

\section{Lempel-Ziv Algorithm}
\label{CLZ}

For the condensation stage of our compression scheme we will utilise 
the basic formalism of the classical universal Lempel--Ziv compression
scheme, transferred to a quantum context. The precise details of 
the Lempel--Ziv method will not be required but we give an 
outline of the method. In fact any other classical scheme that is universal for classical 
i.i.d. sources could be used.

The classical Lempel-Ziv compression scheme \cite{LZ77,covth}
asymptotically compresses the output of an i.i.d. source with unknown
probability distribution to $H$ bits per signal, where $H$ is the
Shannon entropy of the distribution.  It depends upon the fact
that at any time in the decoding process, there is a significant
quantity of data that is known to both sender and receiver.  By
making reference to this data as a shared resource, the sender can
more efficiently transmit further signals from the same source.
The encoder scans the sequence, building up a dictionary of
subsequences in such a way that each new entry in the dictionary
is a 1-bit extension of some previous word.  When the whole
sequence has been parsed in this way, this internal structure of
the dictionary is transmitted as a list of \emph{references};
instead of sending a whole subsequence, its position in the
dictionary is transmitted, plus the single extra bit.  In the limit, these references are
logarithmically shorter than the subsequences they represent.
The decoder reverses this procedure, building up the encoder's
dictionary from the list of references.  The original sequence is
reconstructed simply by concatenating the words of the dictionary.
The Lempel-Ziv code is therefore lossless (i.e. has fidelity 1) and
the compression rate $H$ is achieved as an average value over 
all possible inputs.

Bennett \cite{Benn} showed that it is  possible to implement
any classical algorithm reversibly, with only a polynomial increase in time
and space resources.  We can therefore construct a reversible
version of the Lempel-Ziv algorithm (or any other classical
universal compression algorithm), which may be run on a quantum
computer.

The resulting algorithm treats the orthonormal states of the
computational basis as if they were classical signals. 
For sources which are not diagonal in the computational basis the input
sequence can be regarded as a superposition of ``pseudo-classical''
sequences, each of which is operated upon independently. 
However as we will show below, the
action of the quantum implementation of the Lempel-Ziv algorithm
on such non-diagonal i.i.d. quantum sources is simply
 to condense them to a rate
asymptotically approaching $H$ qubits per signal, where $H\geq S$ is the
Shannon entropy of a suitable probability distribution.  This
feature of $H$ being generally greater than $S$, 
embodies the difficulty arising from a mismatch between the 
eigenbasis of the source and the
computational basis of the computer. As part of our main result 
we will show how this difficulty 
 can be overcome.

In order to help us examine the effect of running the algorithm,
we note that
the quantum implementation (via Bennett's result) 
of any classical deterministic algorithm 
merely enacts a permutation of the set of all strings of computational basis 
states at each step.  That is, each state 
$\ket{e_I}=\ket{e_{i_1}}\otimes \ldots \otimes \ket{e_{i_n}}$ is mapped to some 
$\ket{e_{P(I)}}$, where $P$ is a permutation on the set of sequences $I=i_1 \ldots i_n$; 
no superposition or probabilistic mixing is created.  This action, denoted:
\begin{equation} \label{permutation}
\ket{e_I} \onarrow{C} \ket{e_{P(I)}}
\end{equation}
is much more restricted than that of an arbitrary unitary transformation:
\begin{displaymath}
\ket{e_I} \onarrow{U} \sum_J b_{IJ} \ket{e_J}.
\end{displaymath}
and the restriction will be important for us later (cf Theorem 1 below).

\section{Condensation rate with mismatched bases}
\label{mismatch}

If we knew the eigenbasis of the source density matrix, we could
simply set the computer to use this basis as its computational basis,
and analysis of the output from the algorithm would be relatively
simple.  But since we are aiming to achieve fully universal compression, 
we must assume that we do
not know the source's eigenvectors.

We denote the computational basis by $\basis{C}=\{\ket{e_i} \}$, and denote the 
eigenbasis of the source by $\basis{S}=\{\ket{\lambda_i}\}$.
Our first task is to study the effect of the algorithm on an i.i.d. source
whose eigenbasis $\basis{S}$ does not coincide with the computational basis $\basis{C}$.
We do this by introducing a hypothetical diagonalisation procedure
which has the effect of making the source appear diagonal 
in the computational basis.

\subsection{The Diagonalisation procedure}

Given any orthonormal basis $\basis{}= \{ \ket{i}\}$  with $n$
 states, we define
\begin{equation}
D:\ket{i}\ket{j} \rightarrow \ket{i}\ket{j\oplus i} \;\;\forall
i,j \in \basis{}
\end{equation}
where $\oplus$ denotes addition mod $n$.  This operation is 
commonly used in quantum information processing, particularly in the special
case of $j=0$:
\[ D:\ket{i}\ket{0} \rightarrow \ket{i}\ket{i}\;\;\forall i \in \basis{}
\]
where it serves as a duplication operation.  Note that this only
copies basis states, and not superpositions of them - there is
no conflict with the No-Cloning theorem.  The action of
$D$ on a superposition is an entangled state:
\begin{equation}
\ket{\psi}\ket{0} = \sum_i a_i\ket{i}\ket{0} \rightarrow \sum_i
a_i\ket{i}\ket{i}. \label{D_on_qubit}
\end{equation}
  This is a fatal feature for
those who would like to clone quantum information, but it will be the
key to solving our problem of mismatched bases.

If we apply $D$ to each signal state in the input sequence, we will
produce a duplication (relative to the basis $\basis{}$) 
of this input sequence, entangled with the
original.  If we allow the computer only to operate on the
original, or only on the duplicate, and not to make joint
operations on both, then the state addressed is described by
tracing out one of the systems.
From the RHS of (\ref{D_on_qubit}) we see that the reduced state of either system is
$ \sum_i |a_i|^2 \ketbra{i}{i}$ which is always diagonal in the basis $\basis{}$.

If we take $\basis{}$ to be the computational basis of the computer 
and only address each part of the duplication separately,
 the computer will act on an input that is diagonal in its computational basis.
Note that everything so far is done coherently. Although the
computer is now addressing a mixed state, no measurements have
been carried out, and no information has been lost. 

Now, imagine allowing the computer to address each 
part of the duplication in turn; it carries out
the algorithm on the first sequence, leaving the duplicate
unchanged, then repeats the process on the duplicate, leaving the
first part undisturbed. Alternatively, we can imagine building a
computer twice as large, partitioned into two sides, each of which
works simultaneously on a single copy of the sequence.
Finally apply $D^{-1}$ at each signal position in the resultant state. 

We will show below that
this combined process will leave a final state in the first register
that is identical to the result of simply applying the algorithm
to the given input sequence with no diagonalisation operations being applied.
Since in the alternative process (involving the operation $D$) the algorithm
acts only on states {\em diagonal} in the computational basis,
we can use this equivalence to give a simple derivation of the 
condensation properties that result when the computational basis 
and eigenbasis of the source are not matched.

Consider any classical deterministic algorithm which has been formulated in
a reversible way and implemented on a quantum computer. Thus any step $C$ of
the algorithm is a permutation of the computational basis states, 
in the sense of Eq. (\ref{permutation}).
We wish to prove that the following diagram has a
``pseudo-commutation'' i.e. that the two ways of going
around the loop give the same result:  
\[\begin{CD}
{\ket{e_I}\ket{0}}  @>{C\otimes I}>>    {\ket{e_{P(I)}}\ket{0}} \\
   @ V{\hat{D}} VV                              @ AA {\hat{D}^{-1}} A \\
{\ket{e_I}\ket{e_I}} @>> {C\otimes C} >  {\ket{e_{P(I)}}\ket{e_{P(I)}}}\\
\end{CD}   \]
where $\hat{D}= D^{\otimes n}$, the operation $D$ applied at each position in the sequence.
\paragraph*{}
\begin{theorem_env}
For any unitary operation $C$, the equality $\hat{D}^{-1}(C \otimes C)\hat{D} = (C \otimes I)$ is
satisfied iff $C$ enacts a permutation on the computational basis
states.
\end{theorem_env}

\textbf{Proof:}
If we write $C$ as the unitary operation 
$U: \ket{e_I}\rightarrow \sum_J b_{IJ}\ket{e_J}$, we get:
\begin{align}
 \ket{e_I}\ket{0} & \onarrow{\hat{D}}  \ket{e_I}\ket{e_I}
\onarrow{U\otimes U} \sum_{JK}  b_{IJ} b_{IK} \ket{e_J}\ket{e_K}
\nonumber\\
&\onarrow{\hat{D}^{-1}} \sum_{JK}  b_{IJ} b_{IK}
\ket{e_J}\ket{e_K \ominus e_J}\label{five}
\end{align}
where  $\ominus$ denotes subtraction mod $n$. For this to be
of the form $\sum_J b_{IJ} \ket{e_J}\ket{0}$,it is necessary that 
$\ket{e_K}$ and
$\ket{e_J}$ are always identical i.e. $U$ must map each $\ket{e_I}$ to 
a multiple of a unique $\ket{e_J}$ (and not a superposition): 
\begin{displaymath}
\ket{e_I} \onarrow{U} e^{i\alpha_I} \ket{e_J}.
\end{displaymath}
Finally substituting this form of $U$ into Eq. (\ref{five})
and equating with $U\otimes I \ket{e_I}\ket{0}$ gives $\alpha_I =0$. 
Hence $U$ must be a permutation as claimed. Conversely
if $C$ is a permutation then an easy calculation shows that
$\hat{D}^{-1}(C \otimes C)\hat{D} = (C \otimes I)$ holds. $\QED$

The final state arrived at is the same whether we insert the
diagonalisation operation or not.  The duplication process can
therefore be considered merely a mathematical convenience to save
us from having to directly analyse the action of the condensation algorithm
on non-diagonal inputs.

\subsection{Rate of Condensation}

\label{rate_Bc=Bs} In the special case of $\basis{C}=\basis{S}$,
the computer views the source as emitting classical signals
$\{\ket{e_i}\}$ with probability distribution $\{p_i\}$, where the
$p_i$ are simply the eigenvalues of the source density matrix.
Thus the asymptotic condensation rate is the von
Neumann entropy of the source, $S(\rho)$.  Therefore, in this
special case, the Lempel-Ziv algorithm achieves asymptotically
optimal condensation.

If $\basis{C}\neq \basis{S}$ then by Theorem 1 the condensation 
effect of the algorithm acting directly on the source $\ce$
can be found by considering the reduced state of the duplication of $\ce$. 
The rate achieved will be the same as the condensation rate of a source diagonal
in the computational basis $\basis{C}$, but whose signal probabilities are given 
by the eigenvalues of this reduced density matrix.

If $\lambda_i$ and $\ket{\lambda_i}$ 
are the eigenvalues and eigenvectors of $\ce$ then we can expand 
the eigenvectors in the computational basis $\{ \ket{e_i}\} $ as
\[ \ket{\lambda_i}= \sum_j a_{ij} \ket{e_j} \]
and the source density matrix becomes

\begin{equation}
\rho = \sum_i \lambda_i \ketbra{\lambda_i}{\lambda_i} = \sum_{ijk}
\lambda_i a_{ij} a_{ik}^* \ketbra{e_j}{e_k}
\end{equation}
When we append the ancilla and apply $D$, we get
\begin{equation}
\rho \onarrow{D} \sum_{ijk} \lambda_i a_{ij} a_{ik}^*
\ketbra{e_j}{e_k}_A \otimes \ketbra{e_j}{e_k}_B
\end{equation}
and taking the partial trace over system $B$ gives the reduced
density matrix:
\begin{equation}
\rho^{\prime}  = \sum_j \left( \sum_{i} \lambda_i |a_{ij}|^2 \right)
\ketbra{e_j}{e_j}
\end{equation}

This density matrix, which is diagonal in the computational basis,
describes the distribution of states addressed by the computer 
if the duplication process is carried out.
  We therefore refer to $\rho^{\prime}$ as the
\textit{effective source} density matrix.

The eigenvalues $\mu_j$ of $\rho^{\prime}$ are thus given by
 $\mu_j = \sum_i \lambda_i |a_{ij}|^2$, where the $a_{ij}$ are 
the coefficients in the expansion of source eigenstate $\ket{\lambda}$.  
We may write these coefficients as
$\braket{e_j}{\lambda_i}$, and therefore the $j^{th}$ eigenvalue of $\rho^{\prime}$ 
is:
\[ \sum_i \lambda_i\braket{e_j}{\lambda_i}\braket{\lambda_i}{e_j} =
\bra{e_j} \rho \ket{e_j} \]
i.e. the diagonal matrix elements of $\rho$ when 
$\rho$ is written in the computational basis.
Inserting this expression  into the formula for
Shannon entropy gives:
\begin{equation}
R_c(\rho,\basis{C}) = -\!\!\!\sum_{\ket{e_j} \in \basis{C}}
\bra{e_j}\rho\ket{e_j} \log \bra{e_j}\rho\ket{e_j}
\label{R_c}
\end{equation}
This is the general formula for the asymptotic rate of
condensation for the source $\ce$ 
achieved when we work in an arbitrary computational
basis $\basis{C}$. In the special case
$\basis{C}=\basis{S}$, the formula reduces to $S(\rho)$, as
expected. Furthermore since the matrix $[a_{ij}]$  represents the transition
between two orthonormal bases, it is a unitary matrix and then $[r_{jk}]
\equiv [|a_{kj}|^2]$ is doubly stochastic. Consequently the eigenvalues
$\mu_j = \sum_k r_{jk}\lambda_k$ of $\rho'$ are a doubly stochastic transform
of the eigenvalues of $\rho$, and by a monotonicity theorem for entropy we have
$S(\rho') \geq S(\rho)$. Thus the algorithm with a mismatch of bases 
acts as a standard condensation process
but incurs a loss of optimality of the achieved condensation rate
as quantified by the above formulae.

\section{Truncation}

In the truncation stage we apply the measurements which actually
reduce the physical resources occupied by the quantum information.
The difficulty of truncation is in identifying how many qubits are
``blanks'' and may safely be discarded.  In doing this, we are
effectively estimating the condensation rate achieved, $R_c$,
which depends upon $\rho$ and is therefore not known {\em a priori}.

Before treating the truncation problem itself, we will first
consider a simpler idealised situation. We are presented with a sequence of
$n$ qubits, and told that it is composed of two parts: the first
part is a sequence of $k$ maximally mixed qubits (each in the
state $\frac{I}{2}$), for some integer $0\leq k \leq n$, while the
remaining $n-k$ qubits are in the zero state
$\ket{0}$.  Our task is to determine the value of $k$ as
accurately as possible, in such a manner that, when $n \rightarrow
\infty$ (keeping the fraction $\frac{k}{n}$ constant) the global
fidelity of the state remains arbitrarily high, $F \rightarrow 1$.

We can argue that the state described above resembles the output
from the condensation algorithm. The output state does consist,
approximately, of two distinct substrings: a ``data'' part to be preserved and a
``blank'' part, whose qubits are all in the state $\ket{0}$ 
(except for a small ``tail'' at the end of the data.)  The size of
the ``data'' section is an unknown fraction $\frac{k}{n}$ of the
total length $n$; for large $n$, this fraction is approximately
equal to $R_c(\rho,\basis{C})$, which is independent of $n$.

Bearing in mind these approximations, we can assert that a
solution to the simplified problem will get us most of the way to
a solution of the truncation problem itself.  Afterwards, we will
weaken the assumptions to something more realistic, whilst
preserving the solution.

 Following \cite{SW} we define a projector
\begin{equation}
\Pi_l = I^{1\cdots l} \tensor \ketbra{0^{l+1\cdots
n}}{0^{l+1\cdots n}}
\end{equation}
which acts on a sequence of $n$ qubits, projecting onto the
subspace in which the last $(n-l)$ qubits are in the state
$\ket{0}$. To locate the position $k$ of the boundary
we will develop a strategy that involves applying a sequence
of (suitably smeared) $\Pi_l$'s with decreasing $l$ values.

If we apply this projector to the sequence at some position $l$,
it will (in general) tell us whether the ``boundary'' position $k$
lies to the left or to the right of $l$.  If the projector is to
the right of the boundary then it will certainly project into its
positive subspace, and cause no disturbance.  However, if it lies
to the left of the boundary then it will certainly cause some
disturbance, whichever outcome is obtained. The closer to the
boundary it lies, the greater the disturbance caused.  The
probabilities and magnitudes of disturbance depend upon the number
of maximally mixed qubits to the right of the projector, which we
denote $s$. The projector is then $\Pi_l = \Pi_{k-s}$.

If a projector $\Pi_{k-s}$ projects to its positive subspace
(giving outcome ``1''), then all qubits to the right of
the projector (including those that were maximally mixed) are set
to the zero state. Since $s$ qubits were maximally mixed, the
probability  to project to
$\ket{0}^{\tensor s}$ is simply $2^{-s}$.  
The result of this disturbance is so great that, as
an approximation, we can assign a fidelity of zero to the
resultant state. This is a valid thing to do, since we are only looking
for a lower bound on the fidelity achieved.  We will also need to 
derive an upper
bound on the probability of such an ``error'' in our procedure.

Conversely, if $\Pi_{k-s}$ projects to its perpendicular subspace
(outcome
``0''), the maximally mixed qubits to the right of the projector
are projected \emph{away} from zero.  This is also a disturbance,
although in general a smaller one.  The probability of this is
$(1-2^{-s})$. Using Eq. (\ref{genfid}) the fidelity after the 
disturbance is readily seen to be $(1-2^{-s})$.   The
outcome of the projector tells us that we have
found a position to the left of the boundary, and no more
projections should be made - the process terminates.

Therefore, a projector located $s$ places to the left of the
boundary maintains a fidelity of at least $(1-2^{-s})^2$, and contributes
$2^{-s}$ to the error probability 
(i.e. the probability of not registering the presence of the boundary).

No strategy in which we simply make measurements with $\Pi_l$
projectors can safely give us the information required.  Wherever
we choose to apply the projectors, there can always be 
cases of $l$ values which lie very close to the left of
the boundary value $k$ (i.e. $s$ is very small), and therefore cause a large disturbance.
Admittedly, the probability of such an event is low if $k$ is chosen at random.
 That is, for
most values of $k$ (i.e. most sources) such a strategy would work
well. However, probabilistic success is not good enough for
universal compression, which must give high fidelity for
\emph{all} sources, not just the majority of them.

To avoid this problem we adopt a method of ``smearing''
measurements that was used by Hayashi and Matsumoto \cite{HM,HM2}.
We define a POVM using an equally-weighted average of a
set of projectors $\Pi_l$, each offset by a different amount from a
common basepoint:
\begin{equation}
\bar{\Pi}_l = \frac{1}{Y}\sum_{i=0}^{Y-1} \Pi_{l+i}
\end{equation}
where $Y$, the number of projectors in each POVM ``cluster'', is
some parameter we are free to choose. 
The POVM elements are then $\{\bar{\Pi}_l, I-\bar{\Pi}_l \}$.
Physically we may interpret this POVM as applying a random choice, 
$\Pi_m$, of the projectors $\Pi_l, \ldots ,\Pi_{l+Y}$ 
(chosen with equal prior probabilities $1/Y$) and then forgetting 
the value of $m$. Thus for {\em any} given $k$ value and {\em any} 
choice of $l$, the probability that $\Pi_m$ is close to $k$ 
(and hence causes a large disturbance) is only $O(1/Y)$ which 
can be kept small by choosing $Y$ large enough.

We will use POVMs based on $\bar{\Pi}_{xL}$ for some integer $L$ and
$x\in\{0,1,\ldots,\frac{n}{L}\}$ (i.e. moving in steps of $L$). For
simplicity we assume $L>Y$.  We apply the POVMs in decreasing
order (ie starting with the one furthest to the right).

If all the projectors in the cluster are to the right of the
boundary value $k$, then whichever one is chosen it is sure to
project into its positive subspace, since all the qubits to the
right of that position are certainly zeroes.  Thus no disturbance
is caused, and an outcome of ``1'' is guaranteed.  Given this
outcome, we move to the next lower POVM (ie decrease $n$ by one),
and measure again.  However, these measurements only provide upper
bounds on the value of $k$. To obtain a lower bound on $k$ as
well, we must make a measurement in which some projectors in the
cluster lie to the left of the boundary, in which case some disturbance is inevitably caused.

The expected disturbance depends on $k$ only through the value of
$k\mbox{ (mod }L)$.  It is clear that, since the action of
projectors to the right of the boundary is entirely deterministic
and non-disruptive, we would observe just the same success rate if
the value of $k$ were decreased or increased by $L$ - there would
simply be one more or one less POVM applied to the right.  We can
therefore subsequently ignore the value of $k$ itself, and
consider only $k\mbox{ (mod }L)$, which we denote $K$.  In
principle we must consider each case individually, from $K=0$
through to $K=L-1$, but we find that they divide up into two
classes.

\subsection{Disturbance Bounds}
\noindent
{\bf Fidelity when $K\geq Y$}
\\
When $K\geq Y$, we know that all the projectors in the POVM lie to
the left of the boundary.  The fidelity in this case is therefore
an average over the behaviour of each of these projectors.  The
separation $s$ ranges from $K$, when the leftmost projector is
chosen, to $K-(Y-1)$, when the rightmost projector is chosen.  The
average is therefore:
\begin{equation}
F(K\geq Y) = \frac{1}{Y} \sum_{i=0}^{Y-1} (1-2^{-(K-i)})^2
\label{fk>y_alpha}
\end{equation}
It can easily be seen that the above argument depends only on the
fact that all projectors in the POVM are to the left of the
boundary, and therefore holds for any value of $K \geq Y$.
Expanding the square and neglecting the small squared term we get a simple 
lower bound
\[ F(K\geq Y) \geq 1-\frac{2}{Y} \]
in this case.

\noindent
{\bf Fidelity when $K<Y$}\\
In those cases where $K<Y$, the above argument does not go
through.  The boundary now lies within the ``cluster''.  Some of
the projectors lie to the left of the boundary, but there are
others on the boundary and to the right.  The projectors to the
left can be treated in the same way as above: each has probability
$\frac{1}{Y}$ of being chosen; $s$ ranges from $K$ (for the
leftmost projector) to $1$ (for the projector immediately to the
left of the boundary).  This gives the first term in the formula
below.

When a projector to the right of the boundary is chosen, it will
give the outcome ``1'' with certainty, and we will move to the
next lowest POVM, whose projectors lie around $L$ places further
to the left.  All these projectors are therefore to the left of
the boundary (since $L>Y$).  The fidelity contributed in this case
is the average over that cluster, which is given in the same way
as in the $K\geq Y$ case, but now with $s$ running from $K+L$ to
$K+L-(Y-1)$.  Finally, there are $Y-K$ projectors to the right of
the boundary, each with probability $\frac{1}{Y}$ of being chosen,
which gives the weight on the second term.

Putting this all together gives us:
\begin{align}
F(K<Y) & = \frac{1}{Y} \sum_{s=1}^{K} (1-2^{-s})^2   \nonumber\\
& + \left(\frac{Y-K}{Y}\right)\frac{1}{Y} \sum_{i=0}^{Y-1}
(1-2^{-(K+L-i)})^2 \label{fk<y_alpha}
\end{align}

As in the previous case we expand the squares and neglect the small
square terms:
\begin{align}
&F(K<Y)   \nonumber\\
\geq & \frac{1}{Y}\left(\sum_{s=1}^{K}1\right) -
\frac{1}{Y}\left(\sum_{s=1}^{K}2^{1-s}\right) +
\nonumber\\
+ & \left(\frac{Y-K}{Y^2}\right)\left(\sum_{i=0}^{Y-1}1\right) -
\left(\frac{Y-K}{Y^2}\right)\left(\sum_{i=0}^{Y-1}2^{1-(K+L-i)}\right)
\nonumber\\
= & \frac{K}{Y} - \frac{2}{Y}(1-2^{-K}) \nonumber\\
+ & \left(\frac{Y-K}{Y}\right) -
\left(\frac{Y-K}{Y^2}\right)(2^{1-(K+L)})(2^Y-1) \nonumber
\end{align}

Re-arranging gives:
\[ F(K<Y) \geq 1 - \frac{2}{Y}
\left[(1-2^{-K})-\left(\frac{Y-K}{Y}\right)\left(\frac{2^Y-1}{2^{K+L}}\right)
\right] \] and so
\begin{equation}
F(K<Y) \geq 1- \frac{2}{Y}
\end{equation}
Thus the same lower bound on fidelity applies for all possible
values of $K$, and can be considered a \emph{worst case} value for
fidelity.

Therefore, if we choose $Y$ large enough (recalling that $n$ can be arbitrarily large), 
we can always
guarantee that the fidelity of the truncation process is greater
than $1-\epsilon$ for any given $\epsilon>0$.

\noindent
{\bf Error probability when $K\geq Y$}
\\
We saw above that the probability of a projector $\Pi_l$ at
position $l=k-s$ projecting into its positive subspace, and thus
setting $s$ maximally mixed qubits to zero, is $p_e(s) = 2^{-s}$.
We can therefore write the expected error probability (averaging
over choice of projector) for any value of $K$.

When $K\geq Y$, and the boundary lies outside the cluster, we
simply average $p_e(s)$ over the projectors in the cluster:
\begin{equation}
P_e(K\geq Y) = \frac{1}{Y} \sum_{i=0}^{Y-1} 2^{-(K-i)} =
\frac{1}{Y}\left( \frac{2^Y-1}{2^K}\right)
\end{equation}
Therefore we have an upper bound on the error probability in this
case:
\begin{equation}
P_e(K\geq Y) \leq \frac{1}{Y}
\end{equation}

\noindent
{\bf Error probability when $K<Y$}\\
When the boundary lies within the cluster, an argument similar to
that used for fidelity applies.  The projectors in the
cluster which lie to the left of the boundary each have
probability $\frac{1}{Y}$ of being chosen, so we average over
their contributions.  Additionally, we have a $\frac{Y-K}{Y}$
probability of choosing one of the projectors which lie to the
right of the boundary; if this happens, we move to the next POVM
to the left, and average over the error probabilities for the
projectors in that cluster.  This gives us:
\begin{align}
& P_e(K<Y) = \frac{1}{Y} \sum_{s=1}^{K} p_e(s) \nonumber\\
& + \left(\frac{Y-K}{Y}\right)\frac{1}{Y} \sum_{i=0}^{Y-1}
p_e(K+L-i)
\end{align}

Now expanding the sums and re-arranging:

\begin{align}
& = \frac{1}{Y} \sum_{s=1}^{K} 2^{-s} +
\left(\frac{Y-K}{Y^2}\right)2^{-K-L}\sum_{i=0}^{Y-1} 2^i  \nonumber\\
& = \frac{1}{Y}(1-2^{-K}) +
\left(\frac{Y-K}{Y^2}\right)\frac{2^Y-1}{2^{K+L}}   \nonumber\\
& = \frac{1}{Y}\left[ 1-2^{-K} +
2^{-K}\left(\frac{Y-K}{Y}\right)\left(\frac{2^Y-1}{2^{L}}\right)
\right] \nonumber
\end{align}
so
\begin{equation}
P_e(K<Y) \leq \frac{1}{Y}
\end{equation}
and we have a worst case error probability for all possible values
of $K$. If we can make $Y$ large enough, the probability that we
will erroneously project a qubit to $\ket{0}$ can be made smaller
than any given $\epsilon>0$.

\noindent
{\bf Uncertainty in $k$}\\
We now determine how much information we can expect to obtain from
this procedure.  We continue to apply POVMs until a measurement
gives the ``0'' outcome, saying that we have found a subsequence
that is not all zeroes. When this result is obtained, (and
assuming that we have not made an error), we know the range of
values that $k$ could have.

Given a ``0'' outcome from some POVM, the boundary cannot lie to
the left of this POVM (or else it would certainly have given a
``1'' outcome instead).  Also, it can lie no further to the right
than the rightmost edge of the previous POVM - for if the boundary
\emph{were} there, the previous POVM would already have given a
``0'' outcome (or else made an error).

The position of $k$ is therefore constrained to the $L+Y$
positions between the outermost edges of the last two POVMs.  That
is as much information as we can obtain.

The constraints of high fidelity and low error probability require
us to make $Y$ large.  However, the only constraint on $L$ is the
initial assumption that $L>Y$ (and it is easy to see that removing 
this constraint does not reduce the uncertainty).  
We therefore pick $L=Y+1$, and so
have a final uncertainty in $k$ of the order of $2Y$.  
Thus given any fixed $\epsilon >0$ we can choose $Y=O(\frac{1}{\epsilon})$
(independent of $n$) and hence for all suitably large $n$ we can learn 
the value of $k/n$ with an uncertainty of $2Y/n$ 
(which tends to 0 as $n\rightarrow \infty$) while maintaining
a fidelity of $1-O(\epsilon)$ and error probability $O(\epsilon)$.

\subsection{Weakening the Assumptions}
In the simplification we assumed that each qubit in the first
section was maximally mixed, and every qubit in the second section
was in state $\ket{0}$; in reality, this will not be the case.
These assumptions were not used directly in the argument, but only
to put bounds on the behaviour of the fidelities and projector
probabilities.  We can therefore weaken the assumptions to
something more realistic, whilst preserving the inequalities
derived.

We assume that there exists a number $k$ such that:
\begin{enumerate}
\item A projector $\Pi_{k+s}$  to
the right of that position has a probability to project into its
perpendicular subspace that decreases exponentially with $s$;  the
drop in fidelity when this happens decreases exponentially with
$s$.
\item A projector $\Pi_{k-s}$ to the left of that position has a
probability to project into its positive subspace that decreases
exponentially with $s$; the drop in fidelity when this happens
decreases exponentially with $s$.
\end{enumerate}

These assumptions effectively say
that the components with lengths significantly greater or less
than $R_c$ have exponentially small probability, which we would
expect to be a property of classical i.i.d. sources 
(and even some non-i.i.d. sources)
condensed by a classical universal scheme such as the 
Lempel--Ziv scheme, and hence of our condensation scheme too.  
With these weaker
assumptions in place, the derivations of the bounds on fidelity
and error probability can be repeated.

Thus the procedure defined in this section allows us to truncate the
condensed quantum sequence, removing virtually all the ``blank'' qubits
(except for a small constant number independent of the sequence
length $n$), whilst leaving the fidelity bounded toward unity.
The procedure does not depend upon any prior knowledge of the
source, and can be applied to any sequence for which one can
make the above assumptions - in particular, those sequences which
are produced as output from our condensation algorithm.

\section{Learning the Eigenbasis}

Combining the Condensation and Truncation procedures described
above, we see that, {\em with} knowledge of the eigenbasis of the source
density matrix, but no information about its eigenvalues, optimal
compression to $S(\rho)$ qubits per signal can be attained
asymptotically.  For large enough $n$ the overhead in the achieved
rate, $\frac{Y}{n}$ can be made smaller than any prescribed
$\delta>0$.

However, without knowing the source eigenbasis we can only
compress asymptotically to a rate $R_c$ defined by Eq. (\ref{R_c}).  
Although we do not know the eigenbasis in advance (by
assumption), we now demonstrate that the above procedure can be
used iteratively, to learn the eigenbasis of the source with
arbitrarily small disturbance and hence achieve a fully universal compression
scheme.

Since we can compress the original sequence while maintaining
arbitrarily high fidelity and arbitrarily low probability of
error, we can repeat the process with a different computational
basis.  We can continue iterating the compression-decompression
process as often as we choose; on each iteration the error
probability and fidelity become worse, but still bounded toward
zero and unity respectively
with bounds determined by the value of $Y$ and the number of iterations.
For any given number of iterations we can 
always choose $Y$ large enough to ensure that we meet any
previously specified bounds.

More precisely, for any given $\delta >0$ we may compress the 
source to $S(\rho)+\delta$ qubits per signal as follows (and 
we simultaneously obtain an estimate of the eigenbasis).
Let $d$ be the dimension of the single signal space. The unitary group $U(d)$
viewed as a subset of $C^{d^2}$ inherits the standard euclidean distance
$|U_1 -U_2|$ and it acts transitively on the set of all orthonormal 
bases in the signal space. For any density matrix $\rho$ and $U\in U(d)$
let $H(U,\rho)$ be the condensation rate of Eq. (\ref{R_c}) 
when the mismatch between the computational basis and eigenbasis is 
given by $U$. Thus $H(U,\rho)$ is the Shannon entropy
of the probability distribution $\mu_j$ given by the doubly stochastic transform
 of the eigenvalues $\lambda_i$ of $\rho$:
\[ \mu_j = \sum_i |U_{ij}|^2 \lambda_i \]
and $H(I,\rho)=S(\rho)$ for any $\rho$.

Now for each given $\rho$ and $\delta$ let $\Upsilon(\rho,\delta)$ 
be the largest real number such that
\[ |U-I|<\Upsilon(\rho,\delta) \hspace{5mm} \mbox{implies} 
\hspace{5mm} |S(\rho)-H(U,\rho)|<\delta. \]
Clearly from the definition, $\Upsilon(\rho,\delta)>0$ for all $\rho$ and
$\delta>0$, and if we allow $\rho$ to vary over all density 
matrices then the quantity
\[ \Upsilon(\delta) = \min_\rho \Upsilon(\rho,\delta) \]
will be strictly positive. 

Next note that the group $U(d)$ is compact 
so there exists a {\em finite} mesh $M(\delta)$ of points $\{ V_i\}$
in $U(d)$ with the property that for every $U\in U(d)$ there is a $V_i$ 
with $|U-V_i|<\delta$ (and hence $|U-V_i|<\Upsilon(\rho,\delta)$ 
for any $\rho$). Let $\{ \basis{i}\}$ be the corresponding set of 
bases obtained by applying the transformations of $M(\delta)$ 
to the computational basis. Now given any source 
(with unknown density matrix $\rho$), the above definitions will guarantee 
that the condensation rate $H(V_i,\rho)$ (i.e. obtained by using 
$\basis{i}$ as the computational basis) will have the property that
\[ |S(\rho)-H(V_i,\rho)|<\delta \hspace{1cm}
\mbox{for at least one $i=i_0$} \]
i.e. $H(V_{i_0},\rho)<S(\rho)+\delta$. 

Thus by iteratively compressing and decompressing sequentially 
relative to the finite list of bases $\basis{i}$ we can choose 
the one giving the smallest condensation rate and hence compress 
the source to $S(\rho)+\delta$ qubits per signal. We also learn 
the identity of the minimal basis which then provides an 
estimate of the eigenbasis. Depending on the (fixed, finite) size of 
the set $M(\delta)$ we can choose the fidelity and error bounds 
sufficiently small (i.e. $Y$ and $n$ large enough) 
in each iteration to meet any prescribed bounds
for the total process.

\section{Concluding remarks}
\label{Discussion}

The argument presented here demonstrates that a pure quantum i.i.d.
source can be optimally compressed 
(i.e. compressed asymptotically to its von Neumann entropy)
 with no prior knowledge of the structure of the source.
  The possibility of such universal quantum
compression has also  been recently demonstrated by Hayashi and Matsumoto
\cite{HM} but there are significant differences in the two approaches.
We introduced a diagonalisation procedure that enables any classical algorithm 
to be utilised in a quantum context. For any classical compression algorithm
this gives a  simple method of determining the effect of the resulting 
quantum algorithm on sources that are not diagonal in the computational basis.
This leads to a simplified method of estimating the entropy of an i.i.d. source
in terms of an associated length parameter, while maintaining
high fidelity for sufficiently long blocks of signals.
As a by-product, we also estimate the eigenbasis of the source.

Our diagonalisation procedure may be applied to any classical 
algorithm and it may be interesting to explore its applicability
in other cases (in addition to the Lempel--Ziv algorithm) and even
beyond issues of information compression. The (not universal) 
quantum compression schemes of Schumacher \cite{Schu} 
and Schumacher and Westmoreland \cite{SW} may be viewed as 
similar translations of classical algorithms, but in those cases
knowledge of the source makes it unnecessary to invoke 
the diagonalisation analysis.

Our universal quantum compression scheme was based on a
quantum implementation of the classical Lempel--Ziv algorithm. 
This classical algorithm is known to be applicable to a target 
sequence, such as literary text, that is not produced by an i.i.d. source.
(Indeed the algorithm is widely used in practice for 
compressing computer files). Hence unlike any previously proposed quantum
compression scheme, our scheme is also applicable to such more general
target sequences of quantum states and it would be interesting to explore
its performance for various kinds of non-i.i.d. sources.

Intuitively, the classical Lempel--Ziv algorithm operates by 
building up and continually improving 
a model of the source, based on increasing numbers of
 already received signals, 
which may then be used to reduce the resources needed in 
further transmissions. Our quantum translation merely mimicks 
this procedure in the computational basis and so it would be
especially interesting to investigate whether a ``truly quantum''
extension of the Lempel--Ziv idea exists, in the context of 
quantum information, that is not tied to a particular basis.

\bigskip

\noindent {\Large\bf Acknowledgements}

\noindent RJ is supported by the U.K. Engineering and Physical
Sciences Research Council. SP is supported by the U.K. Engineering
and Physical Sciences Research Council and the U.K. 
Government Communications Head Quarters. We wish to thank A. Winter, 
K. Matsumoto and M. Hayashi
for helpful discussions and comments.


\begin{thebibliography}{10}

\bibitem{js00} Jozsa, R. and Schlienz, J. (2000) ``Distinguishability 
of states and von Neuumann entropy'' {\em Phys. Rev. } 
{\bf A62}, p012301.

\bibitem{LZ77} Ziv, J. and Lempel, A. (1977) ``A universal algorithm for
sequential data compression'' {\em IEEE Trans. Inf. Theory} {\bf
IT-23}, p337

\bibitem{Schu} Schumacher, B. (1995) ``Quantum coding'' {\em Phys.
Rev. }{\bf A51}, p2738

\bibitem{NC} Nielsen, M. and Chuang,. I. (2000) {\em Quantum Computation
and Quantum Information}, Cambridge University Press.

\bibitem{Benn} Bennett, C. H. (1973) ``Logical reversibility of
computation'' {\em IBM J. Res. Devel.} {\bf 17}, p525

\bibitem{SW} Schumacher, B. and Westmoreland, M. D. (2000)
``Indeterminate length quantum coding'' available at 
http://xxx.arXiv.org/abs/quant-ph/0011014

\bibitem{HM} Hayashi, M. and Matsumoto, K. (2002) ``Quantum
universal variable length source coding'' available at
http://xxx.arXiv.org/abs/quant-ph/0202001

\bibitem{HM2} Hayashi, M. and Matsumoto, K. (2002) 
 ``Simple 
construction of quantum
universal variable length source coding'' available at
http://xxx.arXiv.org/abs/quant-ph/0209124

\bibitem{tokyo} A preliminary version of this work was presented at 
the ERATO workshop on quantum information science, 
Tokyo, September 6-8, 2002.

\bibitem{petz} Petz, D. and Mosonyi, M. (2001) 
``Stationary quantum source coding'' 
{\em J. Math. Phys.} {\bf 42}, 4857-4864.

\bibitem{j94} Jozsa, R. (1994) ``Fidelity for mixed quantum states''
{\em J. Mod. Opt.} {\bf 41}, 2314-2323.

\bibitem{u76} Uhlmann, A. (1976) ``The `transition probability' in 
the state space of a *-algebra'' {\em Rep. Math. Phys.} {\bf9}, 273-279.

\bibitem{covth} Cover, T. and Thomas, J. (1991)
 ``Elements of information theory''
Wiley.

\bibitem{js94} Jozsa, R. and Schumacher, B. (1994) 
``A new proof of the quantum noiseless coding theorem''
{\em J. Mod. OPt.} {\bf 41}, 2343-2349.

\bibitem{bcfj} Barnum, H., Caves, C., Fuchs, Ch. and Jozsa, R. (1996)
``General fidelity limit for quantum channels''
{\em Phys. Rev.} A{\bf 54}, 4707.


\bibitem{w99} Winter, A. (1999) 
``Coding theorems for quantum information theory''
PhD thesis chapter 1, University of Bielefeld,Fakult\"{a}t
f\"{u}r Mathematik.\\
(Available at http://xxx.lanl.gov/abs/quant-ph/9907077.)

\bibitem{jhhh} Jozsa, R., Horodecki, M., Horodecki, P. and Horodecki, R. (1998)
``Universal quantuminformation compression''
{\em Phys. Rev. Lett.}{\bf 81}, 1714-1717.

\bibitem{bfg} Braunstein, S., Fuchs, Ch., Gottesman, D. and  Lo, Hoi-Kwong
(1998) ``A quantum analog of Huffman coding'' 
 available at
http://xxx.arXiv.org/abs/quant-ph/9805080.

\end{thebibliography}
\end{document}